\documentclass{cernrep} 
\usepackage{texnames}
\usepackage[T1]{fontenc}
\begin{document}
\title{An Iterative, Dynamically Stabilized~(IDS) Method of Data Unfolding}
 
\author{ Bogdan MALAESCU }

\institute{CERN, CH--1211, Geneva 23, Switzerland}

\maketitle % this produces the title block

\begin{abstract}
We describe an iterative unfolding method for experimental data, making use of a regularization function.
The use of this function allows one to build an improved normalization procedure for Monte Carlo spectra, unbiased by 
the presence of possible new structures in data.
We unfold, in a dynamically stable way, data spectra which can be strongly affected by fluctuations in the 
background subtraction and simultaneously reconstruct structures which were not initially simulated.
\end{abstract}
 
\section{Introduction}
 
\label{SecIntroduction}
Experimental distributions of specific variables in high-energy physics are altered by detector effects.
This can be due to limited acceptance, finite resolution, or other systematic effects
producing a transfer of events between different regions of the spectra.
Provided that they are well controlled experimentally, all these effects can be included 
in the Monte Carlo simulation~(MC) of the detector response, which can be used to correct the data. 

We will not concentrate on the correction of acceptance effects.
It is straightforward to perform it on the distribution corrected for the effects resulting 
in a transfer of events between different bins of the spectrum. 
The detector response is encoded in a transfer matrix connecting the measured and true variables under study. 
However, as the transfer matrix used in the unfolding is obtained from the simulation of a given physical process, one must perform 
background subtraction and data/MC corrections for acceptance effects before the unfolding. 

Several deconvolution methods for data affected by detector effects were described in the
past (see for example~\cite{Hocker:1995kb,Blobel:2002pu,Kondor:1982ah,Lindemann:1995ut,D'Agostini:1994zf,Acton:1993zh}).
We present an unfolding method~(described into detail in~\cite{Malaescu:2009dm}) allowing one to obtain a data
distribution as close as possible to the ``real'' one for rather difficult, yet realistic, examples.
This method is based on the idea that if two conditions are satisfied, namely the MC simulation provides a relatively good 
description of the data and of the detector response, one can use the transfer matrix to compute a matrix of unfolding probabilities. 
If the first condition is not fulfilled one can iteratively improve the transfer matrix.
Our method uses a first step, providing a good result if the difference between data and 
normalized reconstructed MC is relatively small on the entire spectrum.
If this is not the case, one should proceed with a series of iterations.
The regularization of the method is dynamical, coming from the way the data-MC differences are treated in each bin,
at each step of the unfolding method.

This method is to be applied on binned, one dimensional data. 
It can be directly generalized to multidimensional problems.

\section{Important ingredients of the unfolding procedure} 
\label{sec:ImportantIngredients}

\subsection{The regularization function}
\label{Sec:RegF}
We use a regularization function \(f(\Delta x,\sigma,\lambda)\) to dynamically reduce fluctuations
and prevent the transfer of events which could be due to fluctuations, particularly in the subtracted background.
This function provides information on the significance of the absolute deviation \(\Delta x\) between 
data and simulation in a given bin, with respect to the  corresponding error \(\sigma\).
It is a smooth monotone function going from 0, when \(\Delta x = 0\), to 1, when \(\Delta x >> \sigma\).
\(\lambda\) is a scaling factor, used as a regularization parameter. 
As we will see in the following, changing the regularization function used in our method will change the 
way we discriminate between significant deviations and statistical fluctuations.

\begin{figure}[t]
\begin{center}
\includegraphics[width=14cm]{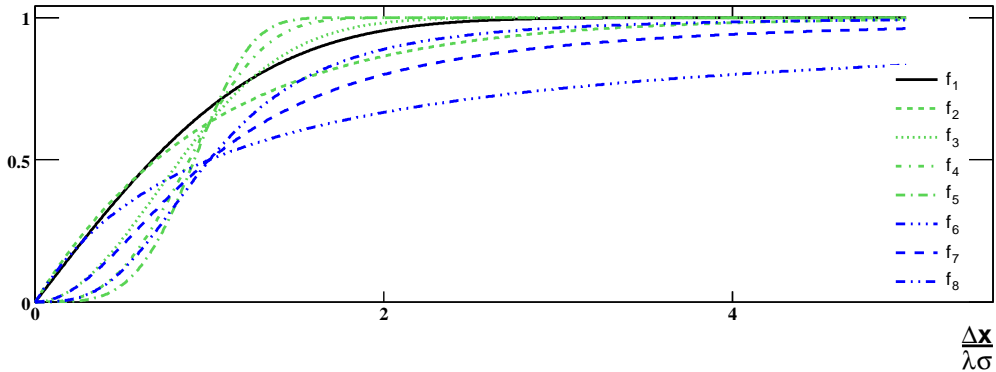}
\caption{Behaviour of the functions \(f_{1..8}\) with respect to \({\Delta x}/{(\lambda \sigma )}\).}
\label{fig:functions}
\end{center}
\end{figure}

For the unfolding procedure, we can consider several functions of the relevant variable \({\Delta x}/{(\lambda \sigma )}\)
(see Fig.~\ref{fig:functions}).
In general, we will use different \(\lambda\) parameters for the regularization function for each componenent of the unfolding procedure described in the following.
We will see however that some of these parameters can be unified~(i.e. assigned identical values) or even dropped~(when a trivial value is assigned to them).

\subsection{The MC normalization in presence of new structures in data} 
\label{Sec:MCnorm}
A rather tricky point is the way the unfolding deals with new structures not considered in the MC generator but present in the data.
These structures are affected by the detector effects, and hence they need to be corrected.
It seems that the Singular Value Decomposition (SVD)~\cite{Hocker:1995kb} and the iterative~\cite{Blobel:2002pu,Kondor:1982ah,D'Agostini:1994zf,Acton:1993zh} 
methods provide a natural way of performing this correction.
However, if the new structures in the data contain a relatively important number of events, they could also
affect the normalization of MC spectra with respect to the data~(needed in the unfolding procedure, as we'll see in the following). 
For the unfolding procedure described here, we introduce a comparison method between data and MC spectra 
which is able to distinguish significant shape differences.
Exploiting the regularization function introduced before, it counts the events in data~(\(N_{d}^{MC}\)) without including those corresponding to significant new structures.
Dividing \(N_{d}^{MC}\) by the number of events in the MC~(\(N_{MC}\)), one obtains the data/MC normalization factor.
This procedure is especially useful when the differences between the two spectra consist of relatively narrow structures. 
Our normalization method allows a meaningful comparison of data and MC spectra and improves the convergence of the algorithm in this case. 
If the differences are widely distributed, they have smaller impact on the normalization factor, and the sensitivity of our method is weaker too.

\subsection{The estimation of remaining fluctuations from background subtraction}
\label{Sec:estBkg}
Experimental spectra are generally obtained after background subtraction;
this operation~(performed before the unfolding) results in an increase in errors for the corresponding data points.
Due to bin-to-bin or correlated fluctuations of the subtracted background, these points can
fluctuate within their errors.
These fluctuations can be important especially on distribution tails or dips, where the signal is weak and the background subtraction
relatively large.
Actually, it is only when the background subtraction produces a large increase in the uncertainties of the data points~(well beyond their original statistical
errors), that these fluctuations become a potential problem.
The problematic regions of the spectrum can be identified even before going to the unfolding.
When computing the corrected distribution, the unfolding procedure has to take into account the size of the experimental errors, 
including those from background subtraction.
At this step we identify large but not significant data-MC deviations.
Not doing so could result in propagation of large fluctuations and uncertainties to more precisely known regions of the spectrum.
Such an effect of the procedure is to be avoided, and we treat this problem carefully.
To our knowledge, none of the previous methods aim at dealing with this second type of problem,
and at distinguishing it from the previous one.

\subsection{Folding and unfolding}
\label{Sec:FoldAndUnf}
In the MC simulation of the detector one can directly determine the number of events 
which were generated in the bin \(j\) and reconstructed in the bin \(i\) (\(A_{ij}\)).
Provided that the transfer matrix \(A\) gives a good description of the detector effects, it is straightforward to compute the corresponding folding and unfolding matrix: 
\(P_{ij} = \frac{A_{ij}}{\sum_{k=1}^{n_{d}}{A_{kj}} } , \tilde{P}_{ij} = \frac{A_{ij}}{\sum_{k=1}^{n_{u}}{A_{ik}} }\).
Here, \(n_{d}\) is the number of bins in data~(and reconstructed MC), while \(n_{u}\) is the one in the unfolding result~(and true MC).
The folding probability matrix, as estimated from the MC simulation, \(P_{ij}\) gives the probability 
for an event generated in the bin \(j\) to be reconstructed in the bin \(i\).
The unfolding probability matrix \(\tilde{P}_{ij}\) corresponds to the probability for the ``source'' 
of an event reconstructed in the bin \(i\) to be situated in the bin \(j\).

The folding matrix describes the detector effects, and one can only rely on the simulation in order to compute it.
The quality of this simulation must be the subject of dedicated studies within the analysis, 
and generally the transfer matrix can be improved before the unfolding.
Systematic errors can be estimated for it and they are propagated to the unfolding result.
The unfolding matrix depends not only on the description of detector effects but also 
on the quality of the model which was used for the true MC distribution.
It is actually this model which can (and will) be iteratively improved, using the comparison 
of the true MC and unfolded distributions.

In order to perform the unfolding, one must first use the iterative procedure described in 
Section~\ref{Sec:MCnorm} to determine the MC normalization coefficient~(\({N_{d}^{MC}}/{N_{MC}}\)).
One can then proceed to the unfolding, where, in the case of identical initial and final binnings, 
the result for \(j\in [1;n_u]\) is given by:
\begin{eqnarray}
 u_j = t_{j}\cdot \frac{N_{d}^{MC}}{N_{MC}} + B^u_j \nonumber 
   + \sum_{k=1}^{n_{d}}{ 
    f \left( \left| \Delta d_k \right| , \tilde{\sigma} d_k , \lambda\right)
    \, \Delta d_k \, \tilde{P}_{kj} +
    \left( 1 - f \left( \left| \Delta d_k \right| , \tilde{\sigma} d_k , \lambda\right) \right)
    \, \Delta d_k \, \delta_{kj}, 
    }
\end{eqnarray}
with \(\Delta d_k = d_k - B^d_k - \frac{N_{d}^{MC}}{N_{MC}} \cdot r_{k}\). 
Here, for a given bin k, \(t_{k}\) is the number of true MC events, while \(\tilde{\sigma} d_k\) is the uncertainty to be used for the comparison of the data~(\(d_k\))
and the reconstructed MC~(\(r_{k}\)).
\(B\) is the (estimated)~vector of the number of events in the data distribution which are associated to a fluctuation in the background subtraction. 
In the case of different binnings for the data and the unfolding, the Kroneker symbol \(\delta\)
must be replaced by a rebinning transformation \(R\).

The first two contributions to the unfolded spectrum are given by the normalized true MC and
the events potentially due to a fluctuation in background subtraction, which we do not transfer 
from one bin to another.
Then one adds the number of events in the data minus the estimated effect from background fluctuations, 
minus the normalized reconstructed MC.
A fraction \(f\) of these events are unfolded using the estimate of the unfolding probability 
matrix \(\tilde{P}\), and the rest are left in the original bin. 
With the description of the regularization functions given in Section \ref{Sec:RegF}, it is 
clear that reducing \(\lambda\) would result in increasing the fraction of unfolded events,
and reducing the fraction of events left in the original bin.
Choosing an appropriate value for this coefficient provides one with a dynamical attenuation of
spurious fluctuations, without reducing the performance of the unfolding itself.

\subsection{The improvement of the unfolding probability matrix}
\label{sec:impMatrix}
As explained in the introduction, if the initial true MC distribution does not contain or badly describes
some structures which are present in the data, one can iteratively improve it, and hence 
the transfer matrix.
This can be done by using a better~(weighted) true MC distribution, with the same folding matrix
describing the physics of the detector, which will yield an improved unfolding matrix.

The improvement is performed for one bin \(j\) at the time, exploiting the difference between an intermediate unfolding result and the true MC~(\(\Delta u_j\)):
\begin{equation}
\label{eq:imprA}
A'_{ij} = A_{ij} + f \left( \left| \Delta u_j \right| , \tilde{\sigma} u_j , \lambda_M\right) \,\Delta u_j  \, P_{ij}
\, \frac{N_{MC}}{N_{d}^{MC}}  \textrm{, for \(i\in \{ 1; N_d\}\) .}
\end{equation}
Here, \(\lambda_M\)~(for modification) stands for the regularization parameter used when modifying the matrix.
Increasing \(\lambda_M\) would reduce the fraction of events in \(\Delta u_j\) used to improve the transfer matrix.

This method allows an efficient improvement of the folding matrix, without introducing spurious fluctuations.
Actually, the amplification of small fluctuations can be prevented at this step of the procedure too.

\section{The iterative unfolding strategy}
\label{Sec:IterUnfStrategy}
In this section we describe a general unfolding strategy, based on the elements presented before.
It works for situations presenting all the difficulties listed before, even in a simplified form,
where some parameters are dropped and the corresponding steps get trivial.
The strategy can be simplified even more, for less complex problems.

One will start with a null estimate of the fluctuations from background subtraction.
A first unfolding, as described in Section \ref{Sec:FoldAndUnf}, is performed, with a relatively large value of \(\lambda = \lambda_L\).
This step will not produce any important transfer of events from the regions with potential 
remaining background fluctuations~(provided that \(\lambda_L\) is large enough), 
while the correction of resolution effects for the new structures in data will be limited too.

At this level one can start the iterations: \\
1) Estimation of the fluctuations in background subtraction \\
An estimate of the fluctuations in background subtraction can be obtained using the procedure
described in Section \ref{Sec:estBkg}.
The parameter \(\lambda_S\) used here must be large enough, in order not to underestimate them.
\(\lambda_S\) can however not be arbitrary large, as this operation must not bias initially
unknown structures, by not allowing their unfolding. \\
2) Improvement of the unfolding probability matrix \\
Using the method described in Section \ref{sec:impMatrix}, one can improve the folding matrix \(A\). 
A parameter \(\lambda_M\), small enough for an efficient improvement of the matrix,
yet large enough not to propagate spurious fluctuations~(if not eliminated at another step), must be used at this step. \\
3) An improved unfolding \\
A parameter \(\lambda_U\) will then be used to perform an unfolding following Section 
\ref{Sec:FoldAndUnf}, exploiting the improvements done at the previous step. 
It must be small enough to provide an efficient unfolding, but yet large enough to avoid
spurious fluctuations (if not eliminated elsewhere).\\
These three steps will be repeated until one gets a good agreement between data and reconstructed MC
plus the estimate of fluctuations in background subtraction.
Another way of proceeding~(providing similar results) could consist in stopping the iterations when the improvement brought by the last one
on the intermediate result is relatively small.
The values of the \(\lambda\) parameters are to be obtained from (realistic)~toy simulations, and some of these can be dropped by using trivial~(null) values for them. 
The needed number of steps can also be estimated using these simulations.

\section{A complex example for the use of the unfolding procedure}
\label{Sec:ComplexEx}
\begin{figure}[t]
\begin{center}
\includegraphics[width=16cm]{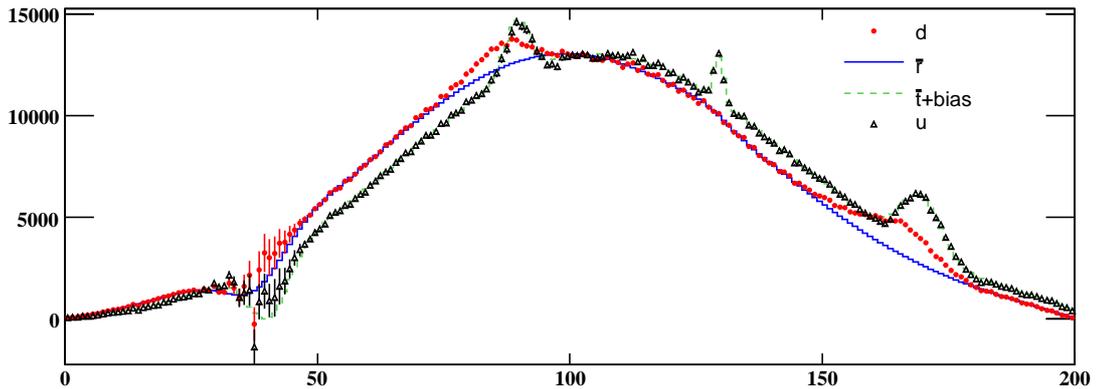}
\caption{The unfolding result after 65 iterations~(u, triangles), compared to the data distribution~(d, filled circles),
the reconstructed MC in the model~(\(\bar{\rm r}\),solid line) and the true MC model plus the new structures~(\(\bar{\rm t}+{\rm bias}\), dashed line).}
\label{fig:toyUnfoldingN1}
\end{center}
\end{figure}

In the following we briefly present a rather complex, yet realistic test, proving the robustness of the method.
It exhibits all the features discussed previously, which are simultaneously taken into account by the unfolding.
For the clarity of the presentation, the structures and dips of the spectrum are separated. 
The structure around the bin 130~(see Fig.~\ref{fig:toyUnfoldingN1}) is present both in data and simulation, while the ones at 90 and 170 are only in data. 
The dip around 40 is affected by large fluctuations due to background subtraction in data. 
All the structures are affected by resolution and a systematic transfer of events, from high to lower bin numbers. 

The first unfolding step was performed with a very large value for \(\lambda_L\) and it corrects all the elements of the spectrum which are 
simulated in the MC, for both kinds of transfer effects~(in spite of the fact that they are relatively important).
The final unfolding result~(after iterations) reconstructs well all the structures in the data model, without introducing
important systematic effects due to the fluctuations in background subtraction~(see Fig.~\ref{fig:toyUnfoldingN1}).
The errors of the unfolding result(s) were estimated using 100 MC toys, with fluctuated data and 
transfer matrix for the unfolding procedure.

Another example for the use of the IDS unfolding method, with less statistics available in the spectra, has been presented in~\cite{KerstinTalk}. 

\section{Conclusions}
We have described a new iterative method of data unfolding, using a dynamical regularization. 
It allows us to treat several problems, like the effects of new structures in data and the large fluctuations from background subtraction, which were not considered before. 
This method has been tested for complex examples, and it was able to treat correctly all the effects mentioned before.

\end{document}